\documentclass[a4paper,preprint,aps,superscriptaddress]{revtex4}
\usepackage{epsfig}

\begin{document}

\title{Energies and radial distributions of $B_s$ mesons - the effect of hypercubic blocking}



%

\collaboration{UKQCD Collaboration}
\noaffiliation
\author{J. Koponen}
\email{jonna.koponen@helsinki.fi}
\affiliation{Department of Physical Sciences and
Helsinki Institute of Physics,
P.O. Box 64, FIN--00014 University of Helsinki,Finland}
\preprint{HIP-2007-13/TH}


%
%

\begin{abstract}
This is a follow-up to our earlier work
for the energies and the charge (vector) and matter (scalar) distributions
for S-wave states in a heavy-light meson, where the heavy quark is static
and the light quark has a mass about that of the strange quark. We study
the radial distributions of higher angular momentum states, namely P- and
D-wave states, using a "fuzzy" static quark.
A new improvement is the use of hypercubic blocking in the time direction,
which effectively constrains the heavy quark to move within a 2$a$ hypercube
($a$ is the lattice spacing).

The calculation is carried out with dynamical fermions on a $16^3 \times 32$
lattice with $a\approx 0.10$~fm generated using the non-perturbatively
improved clover action. The configurations were generated by the UKQCD Collaboration
using lattice action parameters $\beta = 5.2$, $c_\textrm{SW} = 2.0171$
and $\kappa = 0.1350$.

In nature the closest equivalent of this heavy-light system is the $B_s$ meson.
Attempts are now being made to understand these results in terms of the Dirac
equation.
\end{abstract}
\maketitle


\section{Motivation}

There are several advantages in studying a heavy-light system on a lattice.
Our meson is much more simple than in true QCD: one of the quarks is static with
the light quark ``orbiting'' it. This makes it very beneficial for modelling.
On the lattice an abundance of data can be produced, and we know which state we are
measuring -- the physical states can be a mixture of two or more configurations, but
on the lattice this complication is avoided. However, our results on the
heavy-light system can still be compared to the $B_s$ meson experimental
results.

\section{Measurements and lattice parameters}

We have measured both angular and radial excitations of heavy-light mesons,
and not just their energies but also some radial distributions.
Since the heavy quark spin decouples from the game we may label the states as
$\mathrm{L}_{\pm}=\mathrm{L}\pm\frac{1}{2}$, where L is the angular momentum
and $\pm\frac{1}{2}$ is the spin of the light quark.

The measurements were done on a $16^3\!\times\! 32$ lattice with dynamical
clover fermions. We have two degenerate quark flavours with a mass that is close
to the strange quark mass. The lattice configurations were generated by the UKQCD
Collaboration. Some details about the different lattices used in this study
can be found in Table~\ref{LattTable}. Two different levels of fuzzing (2 and 8
iterations of conventional fuzzing) were used in the spatial directions to permit
the extraction of the excited states.

\begin{table}[h]
\centering
 \begin{tabular}{|l|cllll|}
 \hline
 & \# of configs. & $m_q$ & $\kappa$ & $a$ [fm] & $m_{\pi}$ [GeV] \\
 \hline
 DF3 & 160 & $1.1 m_s$  & 0.1350  & 0.110(6)  & 0.73(2)  \\
 DF4 & 119 & $0.6 m_s$  & 0.1355  & 0.104(5)  & 0.53(2)  \\
 \hline
\end{tabular}
\caption{Lattice parameters. These are UKQCD Collaboration's lattices with
$\beta=5.2$ and $C_{\textrm{SW}}=2.0171$.}
\label{LattTable}
\end{table}


\section{2-point correlation function}

\begin{figure}[b]
\centering
\includegraphics[width=0.50\textwidth]{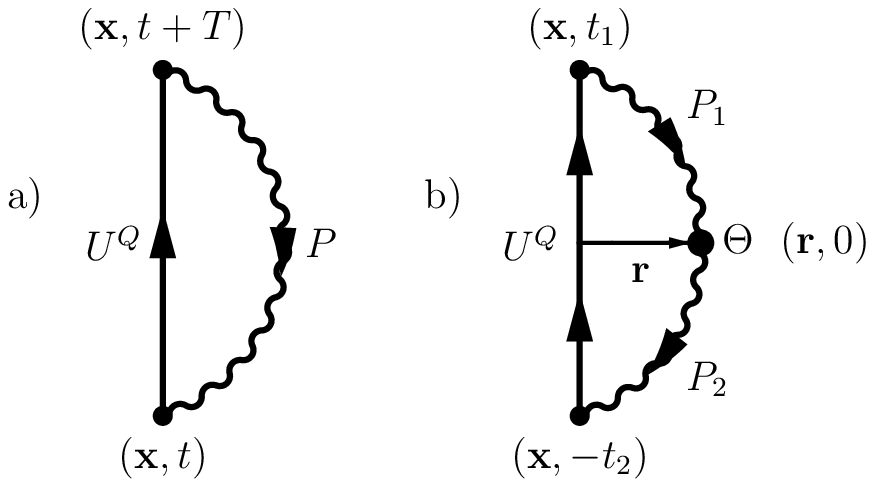}
\caption{Two- and three-point correlation functions}
\label{Fig:Corr}
\end{figure}

The 2-point correlation function (see Figure~\ref{Fig:Corr} a) is defined as
\begin{equation}
\label{2point}
C_2(T)=\langle P_t\Gamma G_q(\mathbf{x},t+T,t)P_{t+T}
\Gamma^{\dag}U^Q(\mathbf{x},t,t+T)\rangle \  ,
\end{equation}
where $U^Q(\mathbf{x},t,t+T)$ is the heavy (infinite mass)-quark propagator
and $G_q(\mathbf{x},t+T,t)$ the light anti-quark propagator. $P_t$
is a linear combination of products of gauge links at time $t$
along paths $P$ and $\Gamma$ defines the spin structure of the operator.
The $\langle ...\rangle$ means the average over the whole lattice.
The energies ($m_i$) and amplitudes ($a_i$) are extracted by fitting the $C_2$
with a sum of exponentials,
\begin{equation}
\label{C2fit}
C_2(T)\approx\sum_{i=1}^{N_{\textrm{max}}}a_{i}\mathrm{e}^{-m_i T},\;
\textrm{where $N_{\textrm{max}}=2\textrm{ -- }4$, $T\leq 14$}.
\end{equation}
Fuzzing indices have been omitted for clarity.

\section{Smeared heavy quark}

\begin{figure}
\centering
\includegraphics[width=0.475\textwidth]{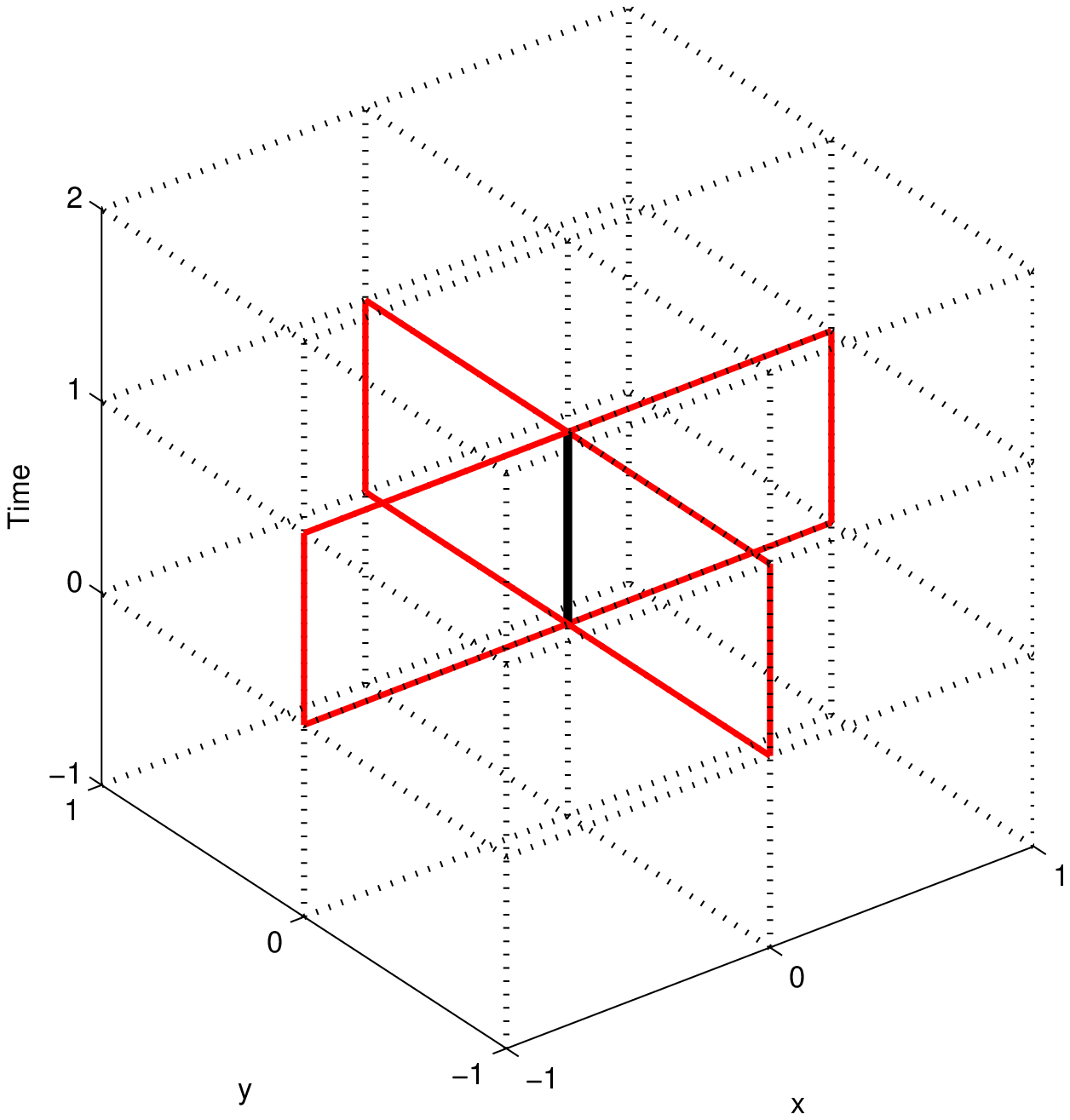}
\includegraphics[width=0.475\textwidth]{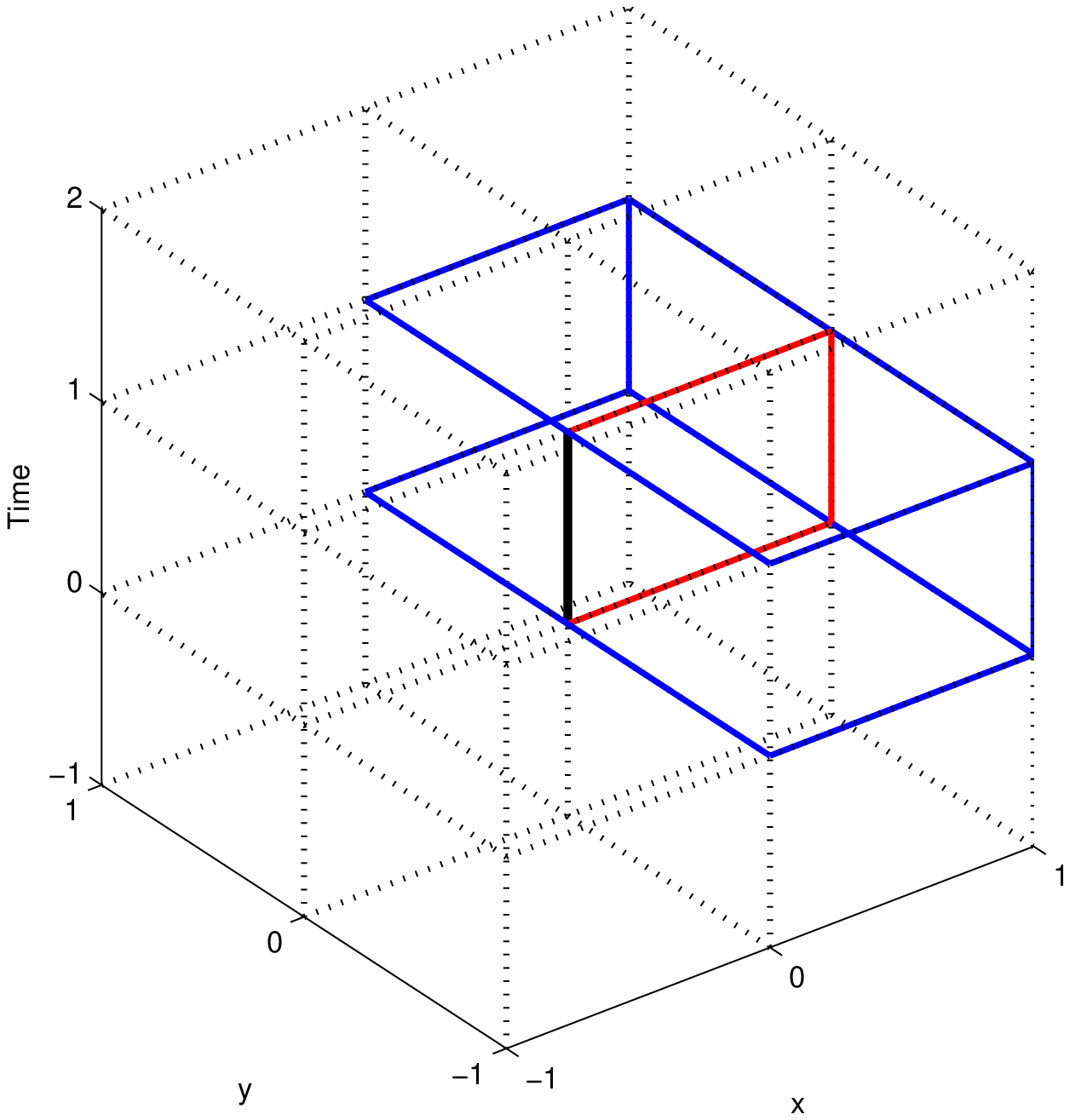}
\caption{APE smearing in the time direction (on the left) and hypercubic
blocking (on the right).}
\label{fig:smear}
\end{figure}

We introduced two types of smearing in the time direction to allow the stationary
quark to move a little, but not too far, from its fixed location. First we
tried APE type smearing, where the original links in the time direction are replaced
by a sum over the six staples that extend in the spatial directions (in
Fig.~\ref{fig:smear} on the left). This smearing is called here ``Sum6'' for short.
To smear the static quark even more we then tried hypercubic blocking, again only
for the links in the time direction (in Fig.~\ref{fig:smear} on the right). Now the
staples (the red ones in Fig.~\ref{fig:smear}) are not constructed of the original,
single links, but from staples (the blue ones in Fig.~\ref{fig:smear}). This allows
the heavy quark to move within a ``hypercube'' (the edges of the ``cube'' are 2$a$
in spatial directions but only one lattice spacing in the time direction). This smearing
is called here ``Hyp'' for short. Smearing the heavy quark was expected to improve
the measurements, particularly radial distributions - which it did, to some extent.

\section{Energy spectrum and spin-orbit splittings}

\begin{figure}
\centering
\includegraphics[height=0.55\textwidth, angle=-90]{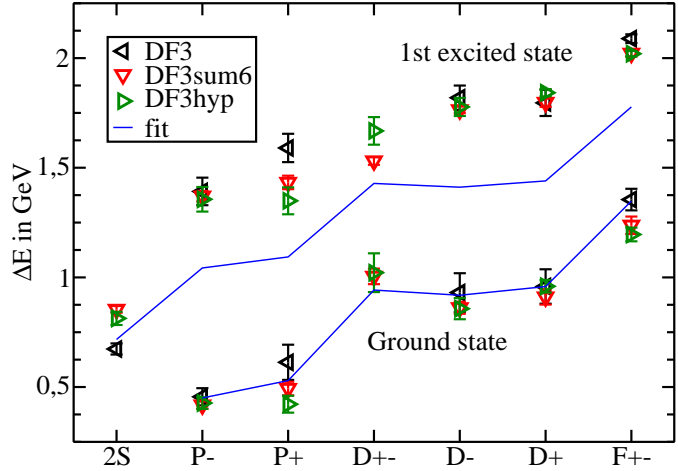}
\caption{Energy spectrum of the heavy-light meson.
Here L$+$($-$) means that the light quark spin couples to angular
momentum L giving the total $j=L\pm 1/2$. 2S is the first radially
excited $L=0$ state. The D$+-$ is a mixture of the D$-$ and D$+$ states, and likewise
for the F$+-$.  Energies are given with respect to the S-wave
ground state (1S). Here $r_0=0.525$~fm was used to convert the energies
to physical units. The error bars shown here contain statistical
errors only. The solid line labelled as ``fit'' is from a model based on the
one-body Dirac equation -- see section \protect\ref{diracmodel} for details.}
\label{fig:espectr}
\end{figure}

The energy spectrum obtained is shown in Fig.~\ref{fig:espectr}. Using different smearing
for the heavy quark does not seem to change the energies too much - except for the
P$+$ state. It is not understood yet why this state should be more sensitive to changes
in the heavy quark than the other states we have considered. The energy of the D$+-$ state
was expected to be near the spin average of the D$-$ and D$+$ energies, but it turned out
to be a poor estimate of this average. Thus it is not clear whether or not the F$+-$ energy
is near the spin average of the two F-wave states, as was hoped. Our earlier results can
be foun in Ref.~\cite{PRD69}.

\begin{figure}
\centering
\includegraphics[height=0.55\textwidth, angle=-90]{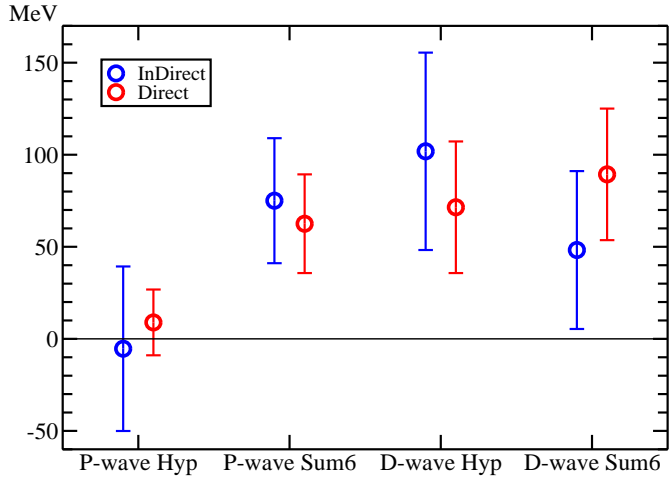}
\caption{The Spin-Orbit splittings of P-wave and D-wave states. The P-wave spin-orbit
splitting seems to be small (could be almost zero), whereas the D-wave spin-orbit splitting
is larger. This is not fully understood yet, but the numbers are still preliminary.
}
\label{fig:SOS}
\end{figure}

One interesting point to note here is that the spin-orbit splitting of the P-wave states
is small, almost zero. We extracted the energy difference of the P$+$ and P$-$ states in
two different ways:
\begin{enumerate}
\item
Indirectly by simply calculating the difference using the energies given by the fits
in Equation~\ref{C2fit}, when the P$+$ and P$-$ data are fitted separately.

\item
Directly by combining the P$+$ and P$-$ data (taking the ratio) and fitting everything
in one go with
\begin{equation}
\frac{\textrm{C}_2(\textrm{P}+)}{\textrm{C}_2(\textrm{P}-)}=
A\exp[-(m_\textrm{P$+$}-m_\textrm{P$-$})T]+B\exp[-m_CT],
\end{equation}
where $A$, $B$ and $m_C$ are fit parameters. $m_\textrm{P$-$}$ and $m_\textrm{P$+$}$
are the energies of the P$-$ and P$+$ ground states, respectively. The energy
difference, rather than the energies themselves, were varied in the fit. The
second exponential contains the corrections from the excited states.
\end{enumerate}
D-wave spin-orbit splitting was also extracted in a similar manner. The results are shown in
Fig.~\ref{fig:SOS}.

%

\section{Radial distributions: 3-point correlation function}

For evaluating the radial distributions of the light quark
a 3-point correlation function shown in Fig.~\ref{Fig:Corr} b is needed. It is defined as
\begin{equation}
C_3(R,T)=\langle \Gamma^{\dag}\, U^Q\, \Gamma\, G_{q1}\, \Theta\,
G_{q2}\rangle.
\end{equation}
This is rather similar to the 2-point correlation function in Eq.~\ref{2point}.
We have now two light quark propagators, $G_{q1}$ and $G_{q2}$, and a
probe $\Theta(R)$ at distance $R$ from the static quark ($\gamma_4$ for the vector
(charge) and $1$ for the scalar (matter) distribution).

\begin{figure}[b]
\centering
 \includegraphics[angle=-90,width=0.495\textwidth]{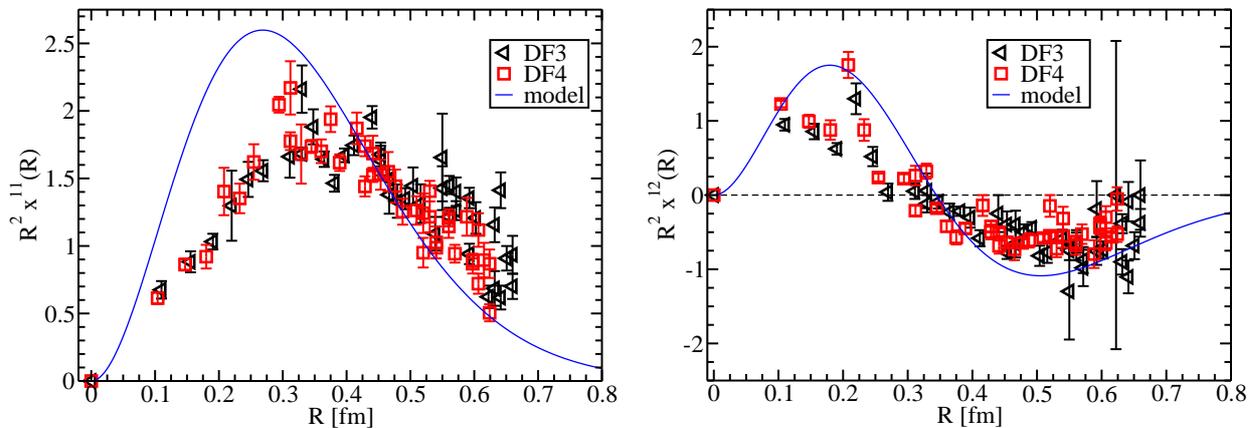}
 \includegraphics[angle=-90,width=0.495\textwidth]{Swave_x12_R2_v2.ps.save}
\caption{On the left:
The S-wave ground state charge distribution. Here we compare two lattices, DF3 and DF4.
The essential difference between the two lattices is the mass of the light quark. However,
the effect of the light quark mass on the distribution seems to be negligible. The label
``model'' on the solid line refers to the model presented in
section~\protect\ref{diracmodel}.
On the right: The S-wave ground state and 1st excited state charge distribution overlap.
Note that we see one node, as expected from the Dirac equation.
}
\label{S11}
\end{figure}

\begin{figure}
\centering
\includegraphics[angle=-90,width=0.495\textwidth]{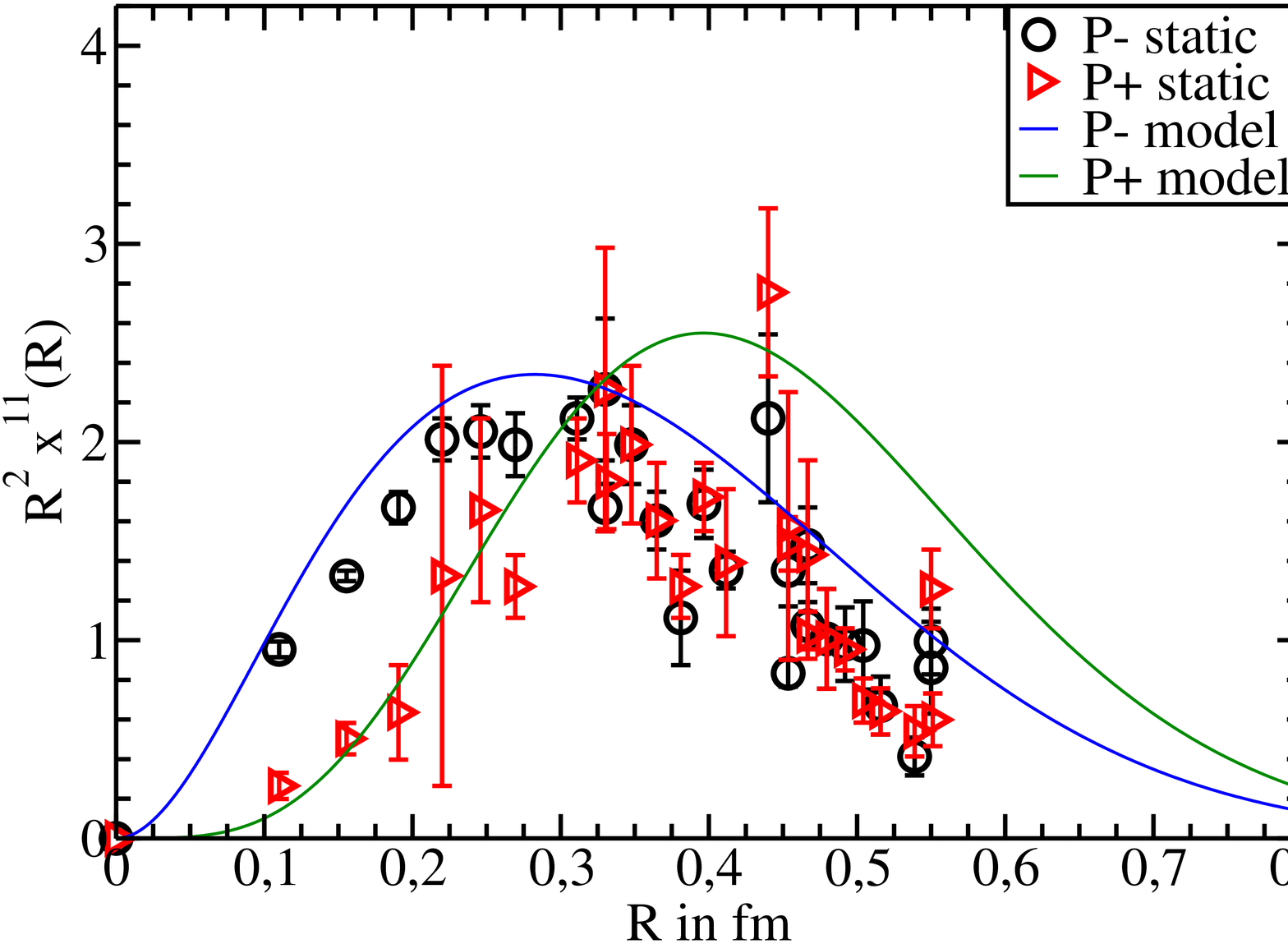}
\includegraphics[angle=-90,width=0.495\textwidth]{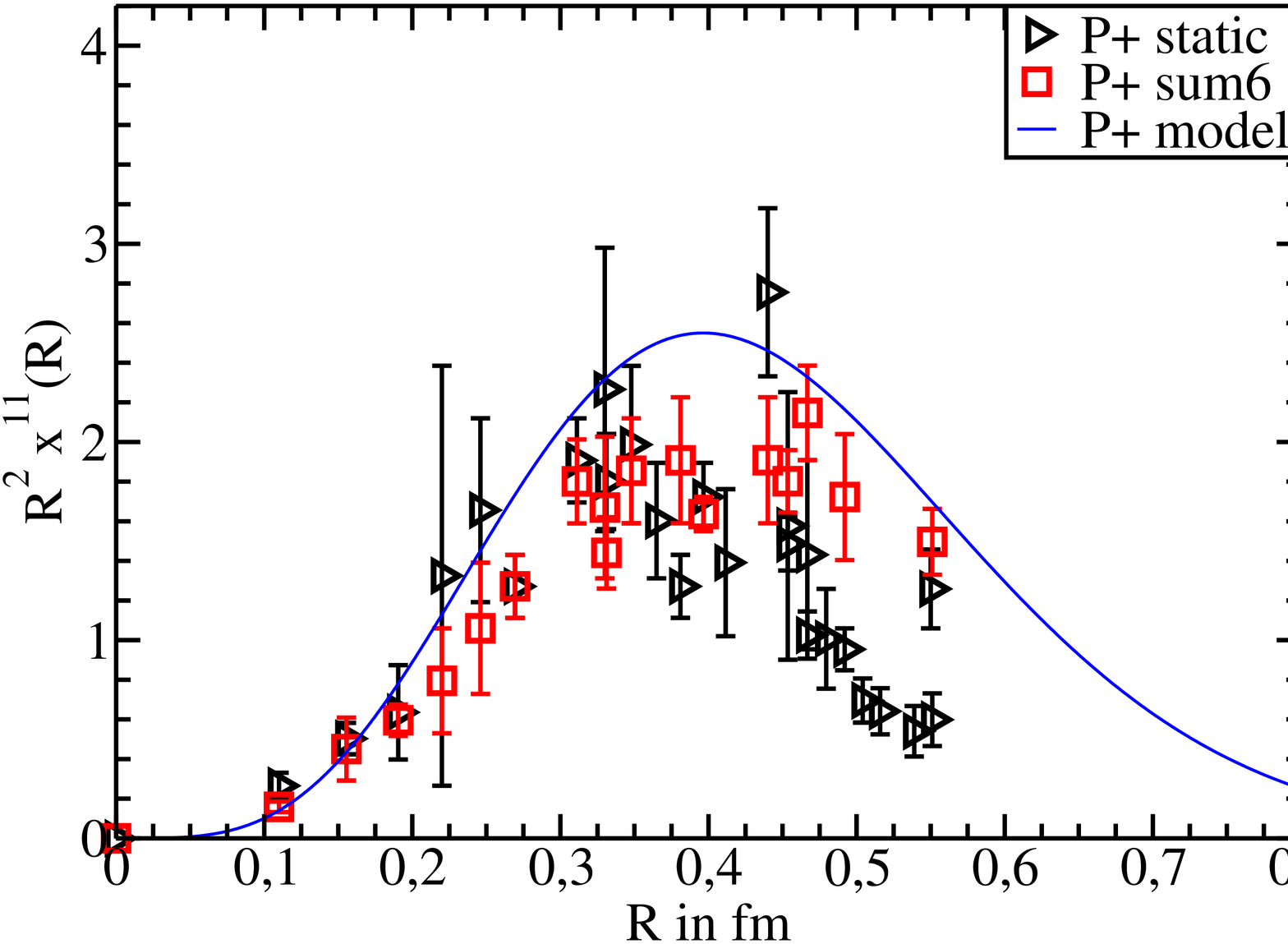}
\caption{On the left:
The P-wave ground state charge distributions. The P$+$ distribution has a peak
a bit further out than the P$-$, which is expected. These measurements were done using
a static heavy quark. The solid lines are predictions from the model in
section~\protect\ref{diracmodel}.
On the right:
The P$+$ ground state charge distribution. Here the results with APE smeared heavy quarks
are compared to the ones with a strictly static quark. Smearing seems to slightly improve the
measurements.
}
\label{P11}
\end{figure}

\begin{figure}
\centering
\includegraphics[angle=-90,width=0.495\textwidth]{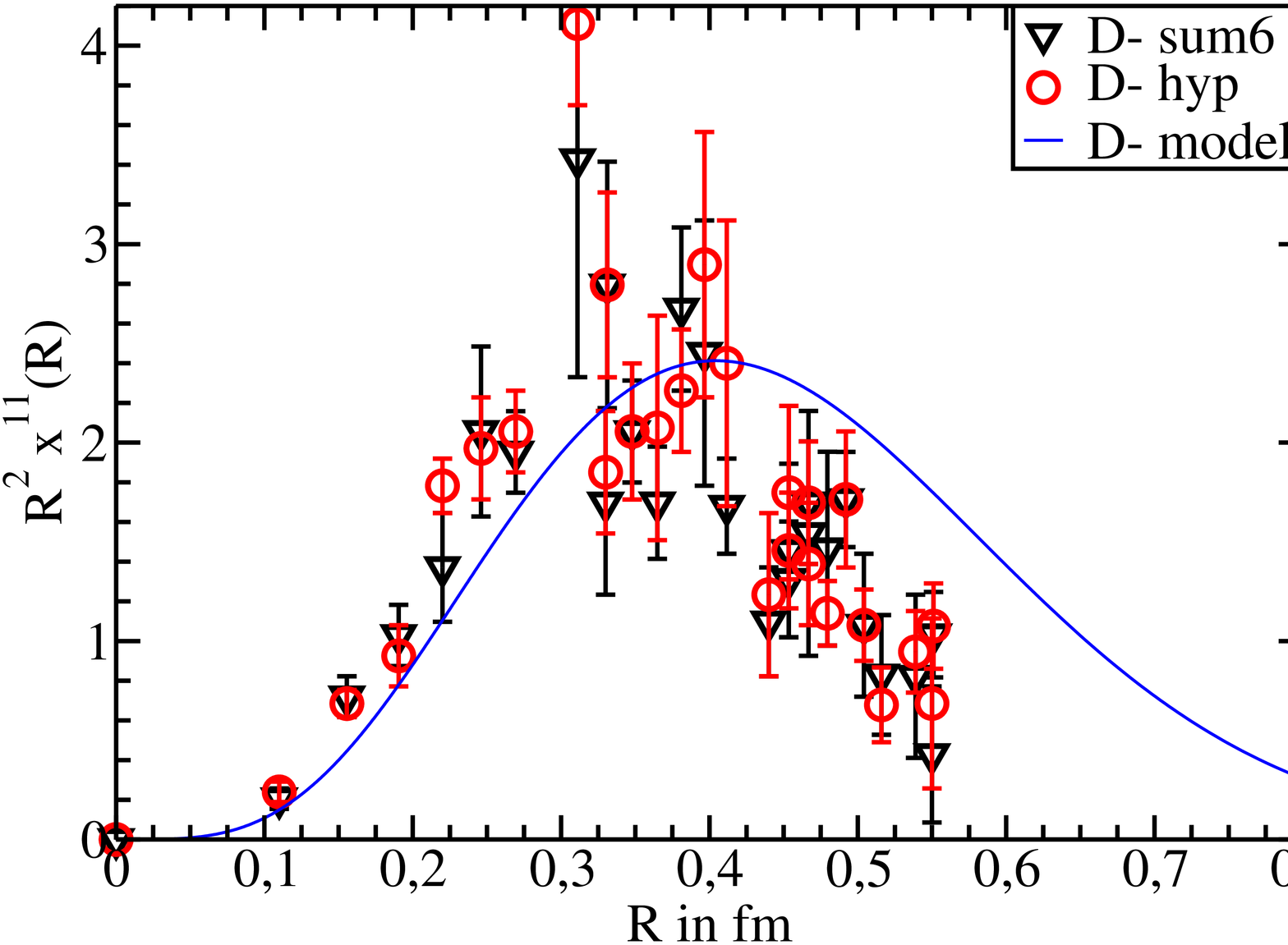}
\includegraphics[angle=-90,width=0.465\textwidth]{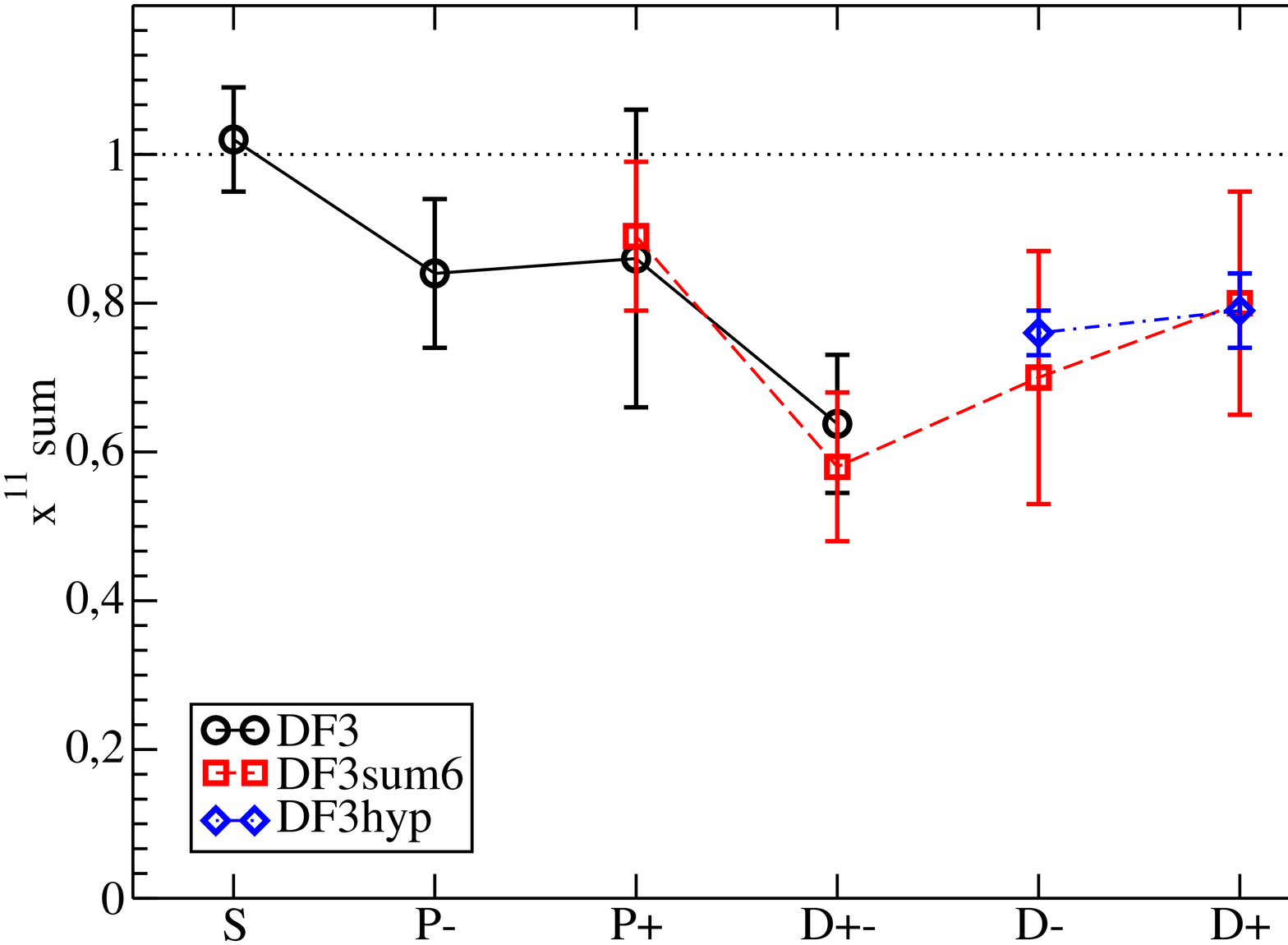}
\caption{On the left:
The D$-$ ground state charge distribution. When the two different smearings of the
heavy quark (APE type smearing, or ``Sum6'' for short, and hypercubic blocking, or ``Hyp'')
are compared, we see that ``Hyp'' gives only a small improvement over the ``Sum6''.
On the right: The charge sumrule.
}
\label{Dmdistr}
\end{figure}

Knowing the energies $m_i$ and the amplitudes $a_i$ from the earlier $C_2$ fit, 
the radial distributions, $x^{ij}(R)$'s, are then extracted by
fitting the $C_3$ with
\begin{equation}
C_3(R,T)\approx\sum_{i,j=1}^{N_{\textrm{max}}}a_{i}\mathrm{e}^{-m_i t_1}%
\; x^{ij}(R)\; \mathrm{e}^{-m_j t_2}a_{j}.
\end{equation}
The results are plotted in Figures~\ref{S11}--\ref{Dmdistr}. The error bars in these
figures show statistical errors only.
See \cite{PRD65} 
for earlier S-wave distribution calculations.
The ``Sum6'' distributions have been published in \cite{Latt05}, 
but the ``Hyp'' results are still preliminary. We are currently trying to improve the
analysis of the D-wave radial distribution data.

\section{Charge Sumrule}

When measuring a radial distribution it is easy to also measure the sumrule by simply summing
over the whole lattice. Our results for the charge sumrule are shown in Figure~\ref{Dmdistr}.
We included the vertex correction $Z_v = 0.7731$, so that with our normalisation the result
should be one. This comes out very nicely for the S-wave, but for the other states we
get the sumrule to be somewhat smaller.


\section{A model based on the Dirac equation}
\label{diracmodel}

A simple model based on the Dirac equation is used to try
to describe the lattice data. Since the mass of the
heavy quark is infinite we have essentially a one-body
problem. The potential in the Dirac equation has a
linearly rising scalar part, $b_{\textrm{sc}} R$, as well as a
vector part $b_{\textrm{vec}} R$. The one gluon exchange potential,
$a_{\textrm{OGE}}\cdot V_{\textrm{OGE}}$, is modified to
\begin{equation}
V_{\textrm{OGE}}(R)\propto \int_0^{\infty}\!\! \mathrm{d}k\, j_0(kR)
\ln^{-1}\frac{k^2+4 m^2_g}{\Lambda^2_{\textrm{QCD}}},
\end{equation}
where $\Lambda_{\textrm{QCD}}=260$~MeV and
the dynamical gluon mass $m_g=290$~MeV (see~\cite{lahde} for details).
The potential also has a scalar term $m\omega L(L+1)$, which is needed
to increase the energy of higher angular momentum states. However, this
is only a small contribution (about 30~MeV for the F-wave).

The solid lines in the radial distribution plots are predictions from
the Dirac model fit with $m = 0.088$~GeV, $a_{\textrm{OGE}} = 0.81$,
$b_{\textrm{sc}} = 1.14$~GeV/fm, $b_{\textrm{vec}} \approx b_{\textrm{sc}}$
and $\omega = 0.028$. These are treated as free parameters with the values obtained
by fitting the ground state energies of P-, D- and F-wave states and the
energy of the first radially excited S-wave state (2S). Note that the
excited state energies in Fig.~\ref{fig:espectr} were not fitted. This
fit was done using the energies obtained with APE smearing (``Sum6''), and
the latest ``Hyp'' data was not used.

\section{Conclusions}

\begin{itemize}
\item
There is an abundance of lattice data, energies and radial distributions,
available.

\item
The spin-orbit splitting is small and supports the symmetry
$b_{\textrm{vec}} = b_{\textrm{sc}}$ as proposed in~\cite{Page}.

\item
The energies and radial distributions of S-, P- and D-wave states can be
qualitatively understood by using a one-body Dirac equation model.
\end{itemize}

\section*{Acknowledgements}

I am grateful to my supervisor A. M. Green and to our collaborator,
Professor C. Michael.
I wish to thank the UKQCD Collaboration for providing the lattice
configurations.
I also wish to thank the Center for Scientific Computing in Espoo,
Finland, for making available resources without which this project could
not have been carried out. The EU grant HPRN-CT-2002-00311 Euridice
and financial support from the Magnus Ehrnrooth foundation are also
gratefully acknowledged.

\end{document}